\documentclass{aastex}
\usepackage{emulateapj5}

\newcommand{\hr}{\mbox{$^h$}}
\renewcommand{\min}{\mbox{$^m$}}
\renewcommand{\deg}{\mbox{$^{\circ}$}}
\newcommand{\lya}{\mbox{Ly$\alpha$}}

\newcommand{\fesc}{\mbox{$f_{\mbox{\tiny esc}}$}}
\newcommand{\fnin}{\mbox{$f_\nu(900 \mbox{ \AA})$}}
\newcommand{\ffif}{\mbox{$f_\nu(1500 \mbox{ \AA})$}}
\newcommand{\fele}{\mbox{$f_\nu(1100 \mbox{ \AA})$}}

\newcommand{\wrst}{\mbox{$W_{\lambda, \mbox{\tiny rest}}$}}
\newcommand{\simgtr}{\, \raisebox{-.2ex}{$\stackrel{>}{\mbox{\tiny $\sim$}}$} \,}

\slugcomment{\it Accepted for publication in the Astrophysical Journal}

\lefthead{Dawson et al.}
\righthead{A Galactic Wind at $z = 5.190$}

\begin{document}

\title{A Galactic Wind at $z = 5.190$\altaffilmark{1}}

\author{
Steve Dawson\altaffilmark{2},
Hyron Spinrad\altaffilmark{2},
Daniel Stern\altaffilmark{3},
Arjun Dey\altaffilmark{4},
Wil van Breugel\altaffilmark{5},
Wim de Vries\altaffilmark{5},
and Michiel Reuland\altaffilmark{5}
}

\altaffiltext{1}{
Based on observations made at the W.M. Keck Observatory, which is operated
as a scientific partnership among the California Institute of Technology,
the University of California and the National Aeronautics and Space
Administration.  The Observatory was made possible by the generous
financial support of the W.M. Keck Foundation.}

\altaffiltext{2}{
Astronomy Department, University of California at Berkeley, Mail Code
3411, Berkeley, CA 94720 USA, email: (sdawson,
spinrad)@astro.berkeley.edu}

\altaffiltext{3}{
Jet Propulsion Laboratory, California Institute of Technology, Mail Stop
169--327, Pasadena, CA 91109 USA, email: stern@zwolfkinder.jpl.nasa.gov}

\altaffiltext{4}{
KPNO/NOAO, 850 N. Cherry Ave., P.O. Box 26732, Tucson, AZ 85726 USA,
email: dey@noao.edu}

\altaffiltext{5}{
Institute of Geophysics and Planetary Physics, Lawrence Livermore National
Laboratory, L--413, P.O.\ Box 808, Livermore, CA 94550 USA, email: (wil,
vries, mreuland)@igpp.ucllnl.org}


\begin{abstract}
We report the serendipitous detection in high--resolution optical
spectroscopy of a strong, asymmetric \lya\ emission line at $z=5.190$.  
The detection was made in a 2.25 hour exposure with the Echelle
Spectrograph and Imager on the Keck II telescope through a spectroscopic
slit of dimensions 1\arcsec\ $\times$ 20\arcsec. The progenitor of the
emission line, J123649.2$+$621539 (hereafter ES1), lies in the Hubble Deep
Field North West Flanking Field where it appears faint and compact,
subtending just 0\farcs3 (FWHM) with $I_{\mbox{\tiny AB}} = 25.4$.  The
ES1 \lya\ line flux of $3.0 \times 10^{-17}$ ergs cm$^{-2}$ s$^{-1}$
corresponds to a luminosity of $9.0 \times 10^{42}$ ergs s$^{-1}$, and the
line profile shows the sharp blue cut--off and broad red wing commonly
observed in star--forming systems and expected for radiative transfer in
an expanding envelope. We find that the \lya\ profile is consistent with a
galaxy--scale outflow with a velocity of $v > 300$ km s$^{-1}$.  This
value is consistent with wind speeds observed in powerful local starbursts
(typically $10^2$ to $10^3$ km s$^{-1}$), and compares favorably to
simulations of the late--stage evolution of \lya\ emission in
star--forming systems. We discuss the implications of this high--redshift
galactic wind for the early history of the evolution of galaxies and the
intergalactic medium, and for the origin of the UV background at $z > 3$.
\end{abstract}



\section{Introduction}
\label{intro}

Following the epoch of recombination, the Universe settled into the
comparatively dormant dark ages, during which the primordial glow had
begun to fade but the first present--day astronomical objects had yet to
form.  This tranquil epoch proved short--lived, however, as the formation
of the first stars and quasars ushered in the first era of cosmological
heating and enrichment at $z < 20$ \citep[e.g.][]{gnedin97, haiman97,
haiman98, ostriker96, valageas99}.  Evidence of these processes in the
form of galaxy--scale outflows is abundant in spectroscopy of the
high--redshift Universe.  Both optical/IR spectra of the $z \sim 3$
Lyman--break population \citep[e.g.][]{pettini01} and optical spectra of
lensed \lya--emitting galaxies at $z > 4$ \citep{frye01}
show metal absorption lines which are blueshifted by
hundreds of km s$^{-1}$ with respect to the stellar rest frame of the
galaxy, and \lya\ emission lines which are shifted similarly redward.
These observations, as well as the characteristic P--Cygni profile of the
\lya\ emission lines (e.g.\ Bunker, Moustakas, \& Davis 2000; Dey et al.\
1997, 1998; Dickinson 1998; Ellis et al.\ 2001, Lowenthal et al.\ 1997; Weymann et al.\ 1998)
paint a coherent picture of optically thick expanding regions surrounding
star--forming galaxies, most naturally driven by the starbursts that
render them visible in the first place \citep[e.g.][and references  
therein]{heckman00}.

We present a direct observation of such an outflow at $z=5.190$ in
high--resolution optical spectroscopy of the serendipitously detected
star--forming galaxy J123649.2$+$621539 (hereafter ES1, for Echelle
Spectrograph and Imager serendipitous detection number one). The sharp blue
cut--off and broad red wing of the ES1 \lya\ emission line are consistent
with the profile expected for the transfer of line radiation in an
expanding envelope \citep[e.g.][]{surdej79}.  The suggested outflow
velocity of $v > 300$ km s$^{-1}$ is in broad agreement with simulations
of the late evolution of \lya\ emission and absorption in star--forming
galaxies \citep[e.g.][]{tenorio99}, and is consistent with observations of
powerful nearby starbursts \citep*[e.g.][]{heckman90}. The spectrum of ES1
therefore presents evidence for both a high star--formation rate and a
high--redshift, starburst--driven galactic wind, fitting well within the
expectations of current models for the early history of galaxy formation.
 
In \S \ref{observation} we discuss the detection of ES1 and we give a
description of the spectrum and the available archival imaging.  In \S
\ref{line} we detail the properties of the ES1 \lya\ emission line, and we
present a model for the emission line profile consistent with the
expanding shell scenario introduced above. We conclude in \S
\ref{discussion} with a discussion of the implications of the evidence for
a strong outflow in ES1 for both the evolution of galaxies and the
intergalactic medium (IGM)  at high redshift, and for the origin of the UV
background at $z > 3$.  Throughout this paper we adopt the currently
favored $\Lambda$--cosmology of $\Omega_{\mbox{\tiny M}} = 0.35$ and
$\Omega_\Lambda = 0.65$, with $H_0 = 65$ km s$^{-1}$ Mpc$^{-1}$
\citep[e.g.][]{riess01}.  At $z=5.190$, such a universe is only 1.10 Gyr
old --- corresponding to a look--back time of 92.1\% of the age of the
Universe --- and an angular size of 1\farcs0 corresponds to 6.31 kpc.

\section{Observation and Data Reduction}
\label{observation}

\begin{figure}
\centering
\epsscale{1.0}
\plotone{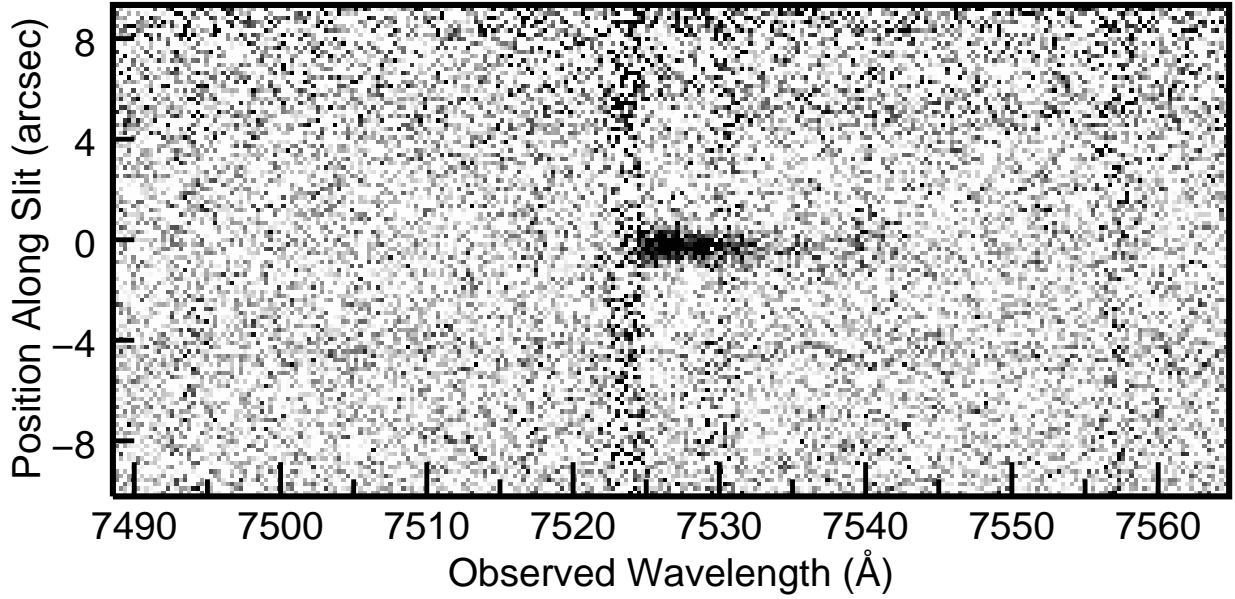}
\caption{
A portion of one order of the ES1 discovery spectrum.  The dispersion axis
is horizontal; the spatial axis is vertical.  The vertical features
flanking the emission line are remnants of the OH and O$_2$ night--sky
emission lines at 7524 \AA\ and 7531 \AA, respectively.  The continuum of
the target galaxy, D16, is barely visible near the slit position
$-4$\arcsec.  See \S \ref{observation} for a description of the
observation.
}
\label{2dspec}
\end{figure}

ES1 was detected in a 2.25 hour exposure made with the Echelle
Spectrograph and Imager \citep[ESI;][]{sheinis00} at the Cassegrain focus
of the Keck II telescope on UT 2001 February 25.  The instrument was
configured in its medium--resolution echellete mode with a spectroscopic
slit of dimensions 1\arcsec\ $\times$ 20\arcsec, yielding a spectral
resolution of $\sim 2$ \AA\ (78 km s$^{-1}$) at 7500 \AA.  The 2.25 hour
exposure was broken into four integrations of 1800 seconds and one
integration of 900 seconds; we performed 3\arcsec\ spatial offsets between
each integration to facilitate the removal of fringing at long
wavelengths.  We used IRAF\footnote{IRAF is distributed by the National
Optical Astronomy Observatories, which are operated by the Association of
Universities for Research in Astronomy, Inc., under cooperative agreement
with the National Science Foundation.} \citep{tody93} to process the
echellogram, following standard slit spectroscopy procedures
\citep*[e.g.][]{massey93}\footnote{{\it A User's Guide to Reducing Slit
Spectra with IRAF}, available online at
http://iraf.noao.edu/iraf/web/docs/spectra.html}.  Some aspects of
treating the ten individual orders of the spectrum were facilitated by the
software package BOGUS\footnote{BOGUS is available online at
http://zwolfkinder.jpl.nasa.gov/$\sim$stern/homepage/bogus.html.}, created
by D. Stern, A.J.  Bunker, and S.A. Stanford.  Wavelength calibrations
were performed in the standard fashion using Xe, HgNe, and CuAr arc lamps;
we employed telluric sky lines to adjust the wavelength zero--point. The
night was near photometric with 0\farcs6 seeing, and we performed flux
calibrations with observations of standard stars from \citet{massey90}
taken with the instrument in the same configuration as the target
observation.  A portion of one order of the discovery spectrum, centered
on the ES1 emission line, is shown in Figure~\ref{2dspec}.

\begin{figure}
\centering
\epsscale{1.0}
\plotone{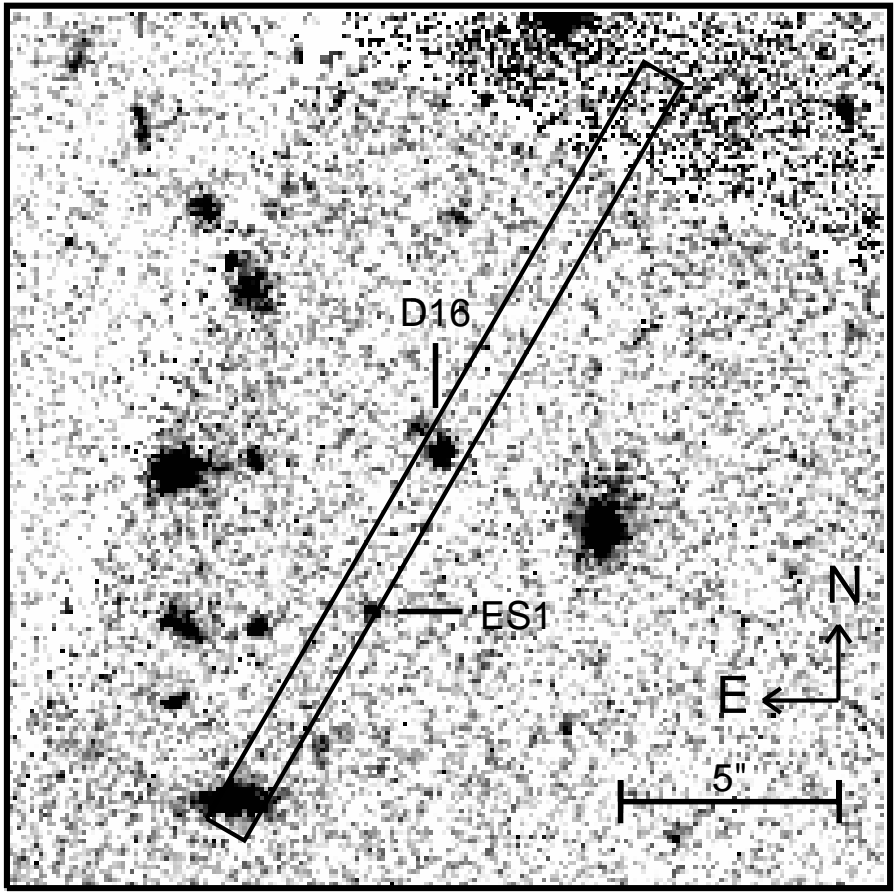}
\caption{
Central region of the {\it HST} $I_{814}$ mosaic of the Hubble Deep Field
North West Flanking Field with a projection of the spectroscopic slit. The
target galaxy (D16) at $\alpha=$12\hr36\min49\farcs0,
$\delta=+$62\deg15\arcmin43\arcsec\ (J2000)  and the serendipitously
detected galaxy (ES1) at $\alpha=$12\hr36\min49\farcs2,
$\delta=+$62\deg15\arcmin39\arcsec\ (J2000)  are indicated.  The panel
measures 20\arcsec\ square and the slit dimensions are 1\arcsec\ $\times$
20\arcsec.  The mosaic and the astrometry therein were provided by
Dickinson (2001, private communication).  The slit position was determined by offsetting from
the set--up star used in the original observation (not shown), thereby
nullifying errors in absolute astrometry.
}
\label{hdfff}
\end{figure}

The target object for this observation was a Lyman--break galaxy at
$z=3.125$ (D16; Steidel 2001, private communication), located in the
Hubble Deep Field North West Flanking Field \citep{williams96}. ES1 was
fortuitously placed on the spectroscopic slit roughly 4\arcsec\ south of
D16, along the parallactic angle of 150$^{\circ}$ (Figure~\ref{hdfff}).  
Upon making the observation, we immediately noticed a strong,
serendipitously detected emission line near 7527 \AA\ whose asymmetric
profile suggested redshifted \lya\ at $z=5.190$.  As the spectrograph
configuration for the discovery spectrum covered only 20 arcsec$^2$ ---
suggesting a surface density of high redshift \lya--emitters roughly 30 to
90 times in excess of current estimates (e.g.\ Cowie \& Hu 1998; Dawson et
al.\ 2001; Stern \& Spinrad 1999; Thompson, Weymann, \& Storrie--Lombardi
2001) --- we consider the detection of ES1 to be highly providential.

Careful inspection of the reduced spectrum failed to reveal emission lines
at other wavelengths. Though the galaxy continuum is faintly detected
redward of the emission line, the observation did not achieve sufficient
signal--to--noise to confirm the presence or absence of interstellar
absorption lines.  The non--detection of high--ionization state emission
lines typically observed in AGN spectra, e.g.\ \ion{N}{5} $\lambda 1240$,
\ion{C}{4} $\lambda1549$, or \ion{He}{2} $\lambda 1640$, strongly suggests
that the \lya\ emission in ES1 is due to the Lyman continuum flux of OB
stars, rather than the hard UV spectrum of an AGN.  For $z = 5.190$, the
nebular emission lines typically associated with star--forming galaxies
are inaccessible to optical spectroscopy.

We were fortunate that ES1 is located in the HDF North West Flanking
Field. Figure~\ref{hdfff} displays a portion of the single--orbit {\it
Hubble Space Telescope} (hereafter {\it HST}) $I_{814}$ image of that
region with a projection of the spectroscopic slit. ES1 is faint and
compact, typical for a very distant galaxy \citep{steidel96}. By running
the source extraction algorithm SExtractor \citep{bertin96} on the
Flanking Field image and employing the conversion from data number to AB
magnitude given in \citet{williams96}, we determine an isophotal magnitude
for ES1 of $I_{\mbox{\tiny AB}} = 25.4 \pm 0.2$. This is challengingly
faint, comparable to the first detected galaxies at $z > 5$, e.g.\ HDF
3--951 with $I_{814} = 25.6$ \citep{spinrad98}.  Moreover, as an
appreciable fraction of the $I$--band radiation for the source is from the
\lya\ emission line, the true continuum magnitude must be still fainter
than that determined from the source extraction.  ES1 appears marginally
resolved, subtending 0\farcs3 (FWHM) on the Flanking Field image, while
stars on the same image have a FWHM near 0\farcs2.  Thus, the physical
size of the emitting region appears to be only $\sim 4$ kpc in diameter.

\begin{figure}
\centering
\plotone{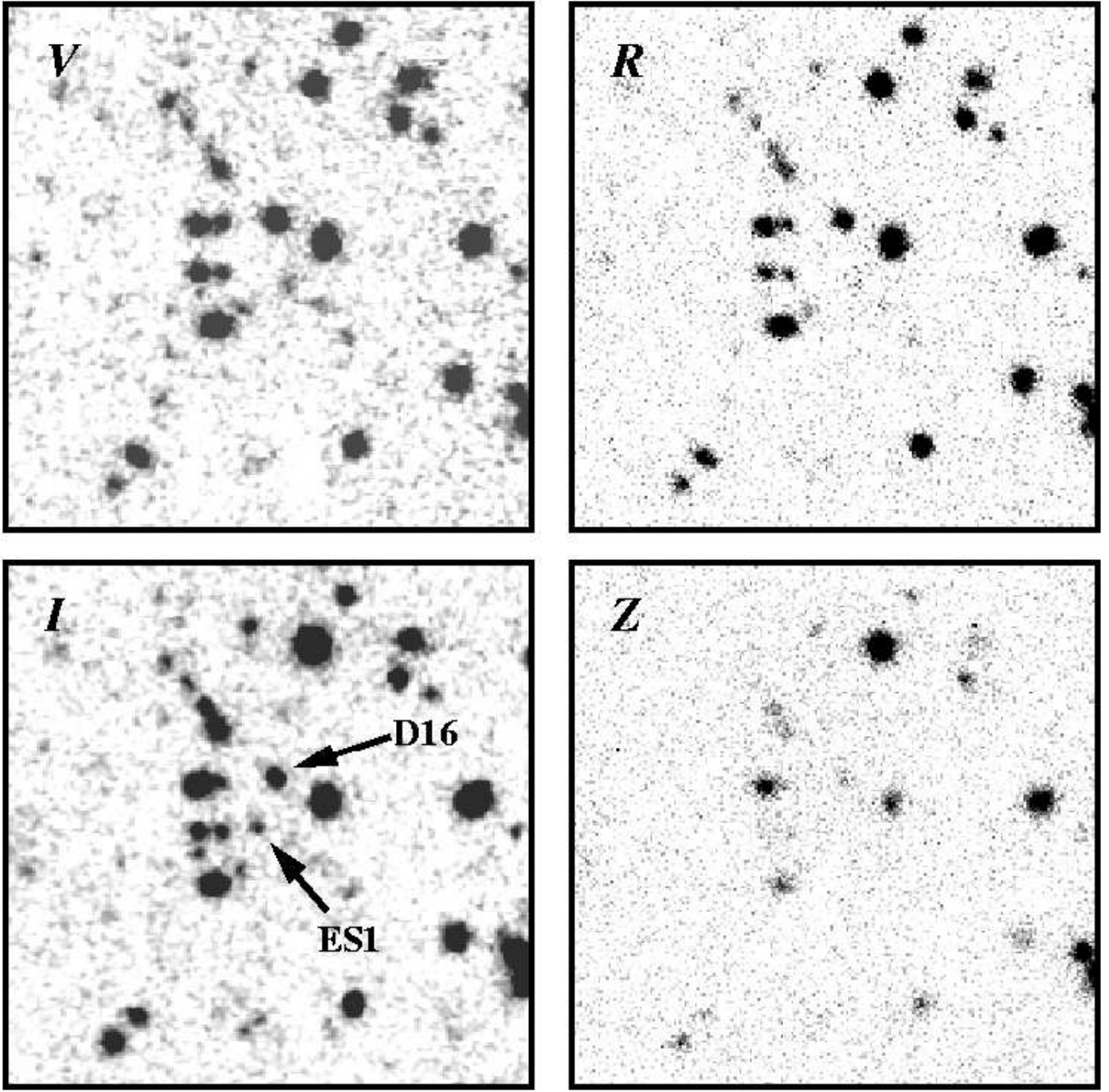}
\caption{
Ground--based supporting imaging for ES1.  The $V$ and $I$ images are from
\citet{barger99};  the $R$ and $z$ images are from our own imaging
campaign of the HDF and its environs \citep[see][]{stern00}.  The fields
are 1\arcmin\ square; North is up and East is to the left.  Notice that
ES1 is not detectable in the $V$ or $R$ bands but is seen in the $I$ band.
This is characteristic of high--redshift galaxies, where intervening
neutral hydrogen (the Lyman forests) severely attenuates the continuum
signal blueward of \lya.  The fact that ES1 is not detectable in the $z$  
band is due only to the shallowness of that image.  We expect the $z$ band  
continuum magnitude for ES1 to be $z > 27.0$; 
the $3\sigma$ limiting
magnitude of the $z$ band image is $z = 25.2$. See Table~\ref{phot}
for a summary of the ground--based photometry.
}
\label{4band}
\end{figure}

We present ground--based $V$, $R$, $I$, and $z$ images of ES1 in
Figure~\ref{4band} and we give the ground--based photometry in
Table~\ref{phot}.  The $V$ and $I$ images are from the Canada France
Hawaii Telescope imaging campaign of \citet{barger99}; the $R$ and $z$
images are from our own Keck imaging campaign of the HDF and its environs
\citep[see][]{stern00}.  The fact that ES1 is not detectable in the $V$ or
$R$ bands is characteristic of high--redshift galaxies, where intervening
neutral hydrogen (the Lyman forests) severely attenuates the continuum
signal blueward of \lya\ \citep{madau95, steidel96, stern99}. The fact
that ES1 is not detectable in the $z$ band is due only to the comparative
shallowness of the $z$ band image.  Based on the model spectrum described
in section \S~\ref{models} and assuming that ES1 has a flat spectrum in
$f_\nu$ at wavelengths longer than the emission line, we expect a
Vega--based continuum magnitude in the $z$ band of $z \simgtr 27.0$.  
This value is almost five times as faint as the $3 \sigma$ limiting magnitude
in that image (see Table~\ref{phot}).

\section{Properties of the \lya\ Emission Line}
\label{line}

\subsection{The Emission Line Luminosity \& Equivalent Width}
\label{linelum}

\begin{figure}
\centering
\epsscale{1.0}
\plotone{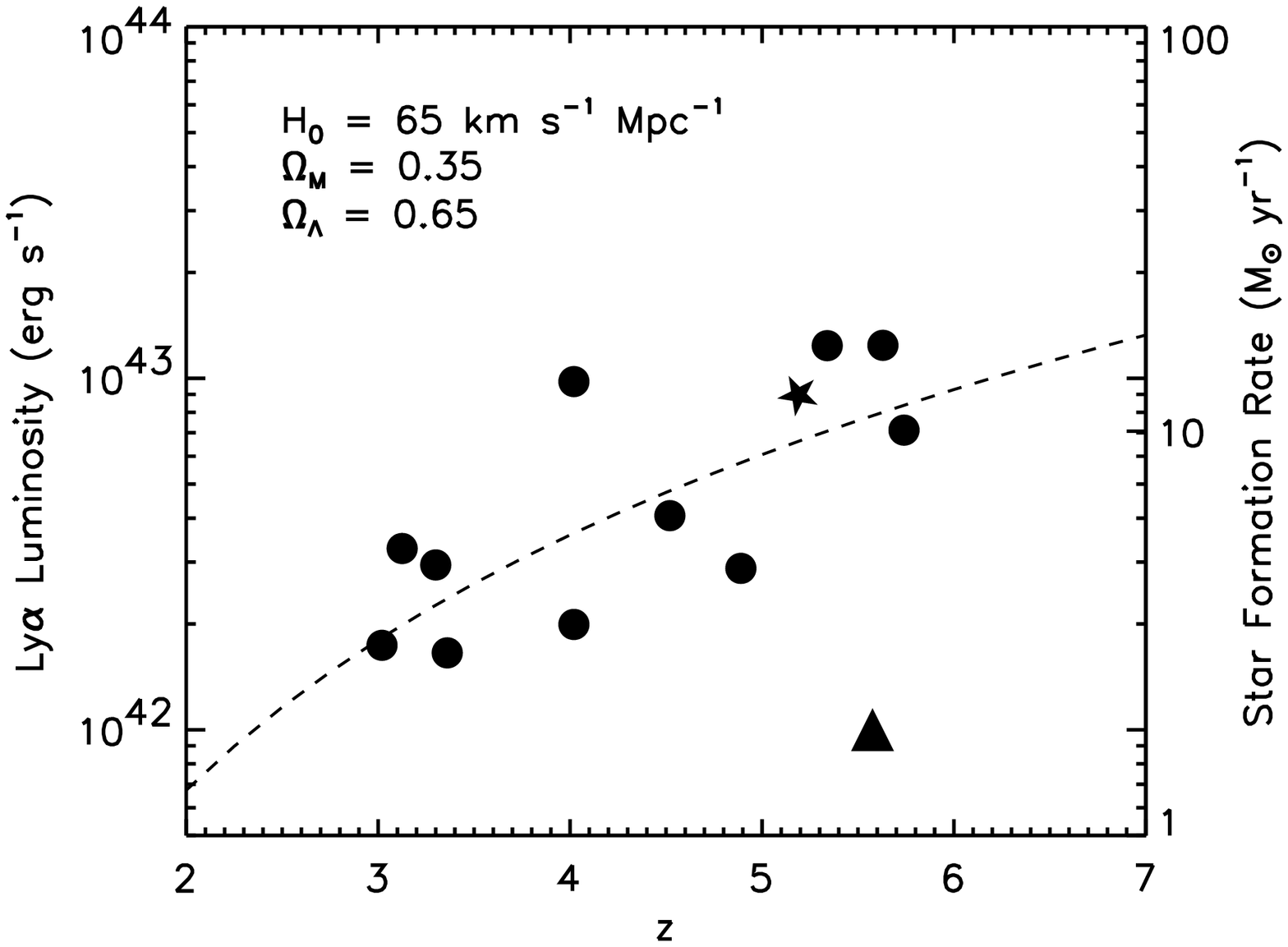}
\caption{
\lya\ emission line luminosity vs.\ redshift for 13 \lya--emitting
galaxies in an $\Omega_{\mbox{\tiny M}} = 0.35$, $\Omega_\Lambda = 0.65$
cosmology.  ES1 is indicated with a star.  The SFR scale has been adopted
from \citet{dey98}.  The sample of galaxies represented by circles 
was compiled from
\citet{dawson01}, \citet{dey98}, \citet*{hu99}, \citet{manning00},
\citet{rhoads00}, \citet{spinrad99}, Stanford (2001, private communication), and
\citet{stern99}.  
The galaxy represented by the triangle is from \citet{ellis01}.
The dashed line indicates the limiting sensitivity to
line flux in the Large Area Lyman Alpha Survey \citep[$\sim 2 \times
10^{-17}$ erg cm$^{-2}$ s$^{-1}$;][]{rhoads00}.
} 
\label{lums}
\end{figure}

In the fashionable cosmology $\Omega{\mbox{\tiny M}} = 0.35$,
$\Omega_\Lambda = 0.65$, the ES1 \lya\ flux of $3.0 \times 10^{-17}$ ergs
cm$^{-2}$ s$^{-1}$ corresponds to a line luminosity of $9.0 \times
10^{42}$ ergs s$^{-1}$.  For the prescription given in \citet{dey98}
assuming no dust aborption and negligible extinction, this luminosity
corresponds to a star formation rate (SFR) of $\sim 10$ M$_{\odot}$
yr$^{-1}$.  As is evident in Figure~\ref{lums}, these values are typical
of \lya--emitting galaxies at high redshift ($z > 3$) discovered
serendipitously \citep[e.g.][]{dawson01, dey98, manning00, spinrad99,
stern99} or in narrow band surveys \citep[e.g.][]{hu99, rhoads00}.
Of course, such galaxies may represent rare systems at the 
high--luminosity tip of an unexplored underlying population, and
the observed \lya\ luminosities may accordingly be governed by a selection
effect.  This point is borne out by the faint system discovered in the 
blind spectroscopic survey of well--constrained lensing clusters
\citep{ellis01}, which is at a higher redshift ($z = 5.576$) and
is an order of magnitude less luminous in \lya\ than ES1.

We estimate a rest--frame equivalent width for the emission line of $\wrst
= 120 \pm 40$ \AA\ based on the emission line model described in the
following section.  This result should be treated with a degree of
circumspection, however. The foremost caveat is that both the total line
flux and the continuum level which enter the calculation were obtained
from the model spectrum before attenuation by neutral hydrogen absorption.
In this manner we attempted to circumvent the ambiguity in deriving an
equivalent width from a strongly P--Cygni line profile coupled with a
pronounced continuum break, essentially arriving at a theoretical
equivalent width from the spectrum as it would appear if not truncated by
foreground absorption. As a second caveat, the continuum redward of the
emission line is not well--detected, with a significance of far less than
$1 \sigma$.  Together, we expect these two effects to cause an
over-estimation of \wrst. However, this tendency toward over--estimation
may be offset by the fact that this is a serendipitous detection, so ES1
was not well centered in the spectroscopic slit (see Figure~\ref{hdfff}).
As such, we have been conservative in our estimate of the uncertainty in
\wrst, which includes sky noise, the uncertainty in the continuum level
redward of the emission line in the extracted spectrum, and the fit errors
in the model spectrum. Even so, with $\wrst = 120 \pm 40$ \AA, ES1 figures
in the top 2\% of Lyman--break galaxies with \lya\ in emission in the
near--complete continuum--selected sample of \citet{steidel00}.

\subsection{Modeling the Emission Line}
\label{models}

The asymmetric \lya\ emission lines commonly observed in high--redshift
starburst galaxies are generally ascribed to the interaction of Lyman
continuum photons generated by newborn OB associations with a
galaxy--scale expanding shell of neutral hydrogen.  For a sufficiently
massive starburst, the hot ionized gas created in the vicinity of the
stars vents into the halo of the galaxy, where it sweeps up neutral
hydrogen into an optically thick shell.  Recombination in the ionized gas
converts Lyman continuum photons escaping from the surface of the hot
stars into line photons.  Then, from the vantage of an observer, the near
side of the expanding shell absorbs photons on the blue side of the
resonant \lya\ emission line, causing a flux decrement on what would
otherwise be the blue wing of the \lya\ emission.  The far side of the
shell back--scatters \lya\ photons into the observer's line--of--sight; as
these photons are offset redward by hundreds of km s$^{-1}$ from both the
rest frame of the galaxy and the approaching side of the neutral shell,
they escape the galaxy and impose a pronounced red wing on the emission
line profile.  The net effect is to create the P--Cygni profile ubiquitous
in observations of expanding shells.

The comparatively high signal--to--noise of the ES1 \lya\ detection
created the opportunity for probing this emission line structure with a
simple model. In accordance with the scenario described above, we fit the
\lya\ emission feature with three components: (1)  a comparatively large
amplitude, narrow Gaussian intended to model line radiation generated by
recombination in the ionized hydrogen;  (2) a small amplitude, broad
Gaussian intended to model the red wing of line photons back--scattered
off the far side of the expanding shell; and (3) a Voigt absorption
profile intended to model the blue decrement caused by the absorption of
line photons by the near side of the shell.  We modeled the weak continuum
with a constant ($f_\lambda \propto \lambda^0$) baseline.

To account for the continuum decrement caused by line blanketing in the
\lya\ forest, we attenuated the model spectrum by a transmission profile
adopted from \citet{madau95}, where the optical depth due to the combined
effect of many \lya\ absorption lines is given as
\begin{equation} 
\tau = 0.0036 \left ( \frac{\lambda}{\lambda_\alpha} \right )^{3.46} \; , 
\end{equation} 
with $\lambda_\alpha = 1216$ \AA.
At high redshifts ($z > 4.5$), absorption by metal systems makes a
non--negligible contribution to cosmic opacity.  Hence, we accounted for
the combined effect of metal lines with the additional factor also given
in \citet{madau95},
\begin{equation} 
\tau = 0.0017 \left ( \frac{\lambda}{\lambda_\alpha} \right )^{1.68} \; .  
\end{equation}
Neither Lyman series line blanketing nor continuum absorption from neutral
hydrogen were included, as both these effects fall shortward of the
wavelength range of interest.

\begin{figure}
\centering
\epsscale{1.0}
\plotone{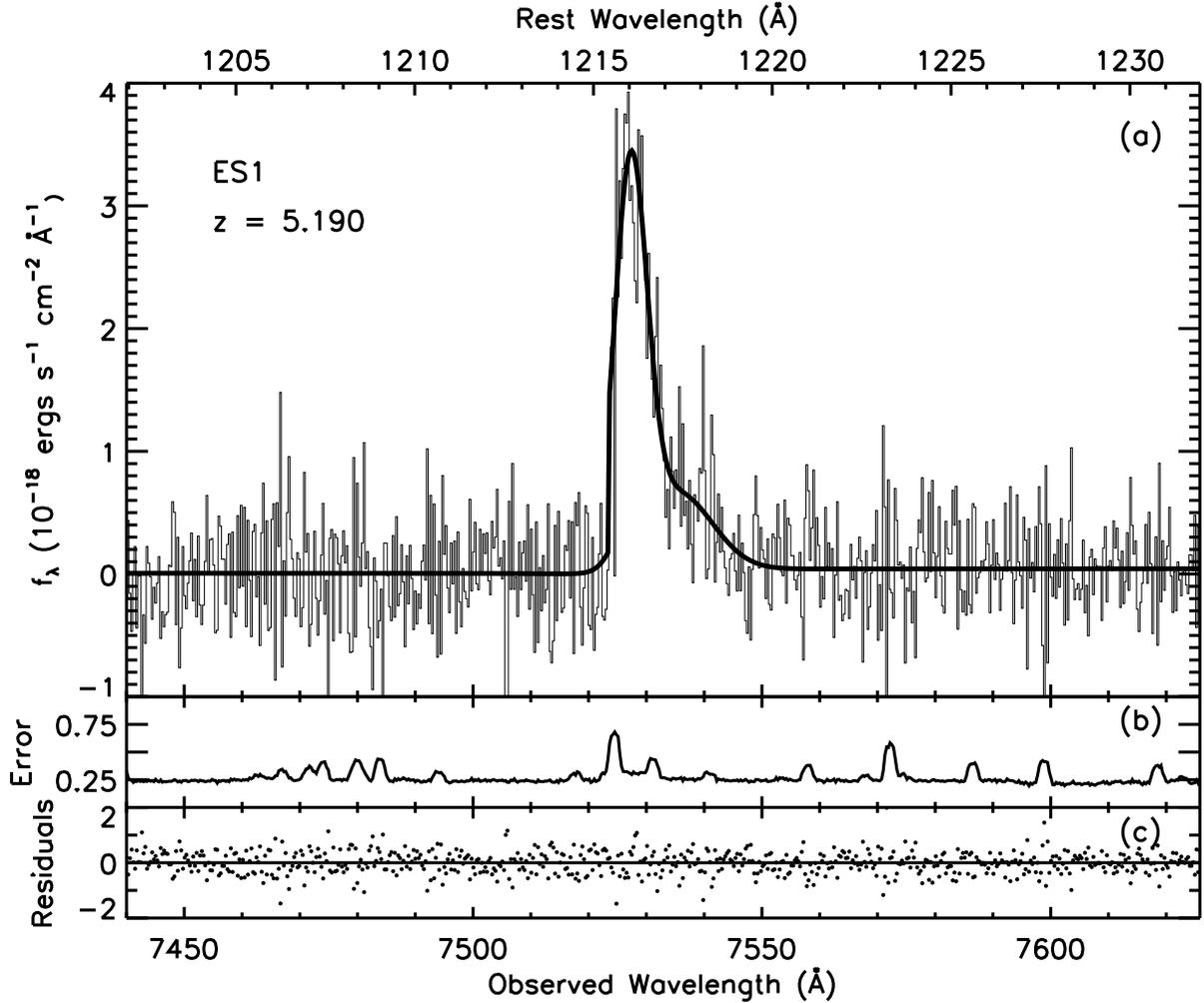}
\caption{
(a) The minimum--$\chi^2$ fit to the ES1 \lya\ emission line.  The line
profile is the sum of (1) a narrow (280 km s$^{-1}$ FWHM) central Gaussian
intended to model recombination in the hot ionized gas of the starburst,
(2) a broad (560 km s$^{-1}$ FWHM) Gaussian redshifted by 320 km s$^{-1}$
from the central component intended to model back--scattering off of the
far side of an expanding shell, and (3) a broad (800 km s$^{-1}$ FWHM)  
Voigt absorption component blueshifted by 360 km s$^{-1}$ from the central
component intended to model absorption by the near side of the expanding
shell.  The model spectrum was attenuated by the model of the \lya\ forest
presented by \citet{madau95}.  (b) The error per pixel in the same flux
units and over the same wavelength range.  
For background--limited observations of faint
objects in this region of wavelength space, 
night--sky emission lines are the dominant source of noise.  (c)
The model--fit minus the data in the same flux
units and over the same wavelength range.  The even distribution of the residuals
demonstrates a lack of systematic error in the model.
}
\label{modspec}
\end{figure}

Figure~\ref{modspec} shows the minimum--$\chi^2$ model resulting when the
centroids, amplitudes, and widths of each of the model components are left
unconstrained.  We weighted the $\chi^2$--fit by the error spectrum
shown in the figure; this had the effect of diminishing
the contribution to the fit by pixels which fall on 
OH and O$_2$ night sky emission lines (which are
the dominant source of error in low signal--to--noise spectra
over the wavelengths of interest).  We note
that the residuals are distributed evenly about zero, demonstrating an
encouraging lack of systematic error in our model.  Furthermore, when just
the components intended to model the expanding shell were examined (the
red, broad Gaussian and the blue Voigt--profile absorption), we found the
resulting profile to be in excellent qualitative agreement with the
P--Cygni line profiles expected for the transfer of line photons in
expanding envelopes \citep[e.g.\ see][and references therein]{surdej79}.

Table~\ref{fittab} summarizes the minimum--$\chi^2$ model parameters.  
This best--fit model yields a redshift for the central component of the
ES1 emission line of $z = 5.190 \pm 0.001$.  Most strikingly, to fit the
red wing of the emission line, the model demands that the broad emission
component be displaced by 320 km s$^{-1}$ from the central emission
component.  Owing to the strong signal of the \lya\ forest at such a high
redshift, only a small blue absorption component is required.  Still,
though of minor amplitude, the absorption component is displaced by fully
$360$ km s$^{-1}$ from the central emission component.  These displacement
velocities are in broad agreement with those predicted by
\citet{tenorio99} for the late stages of the evolution of the \lya\
profile of a star--forming galaxy.  Moreover, though these values somewhat
exceed the $\sim 10^2$ km s$^{-1}$ winds typically observed in nearby
starbursting dwarf galaxies \citep[e.g.][]{martin98}, they compare
favorably with the $10^2$ -- $10^3$ km s$^{-1}$ outflows observed in more
powerful local starbursts \citep[][]{heckman90}.

\begin{figure}
\centering
\epsscale{1.0}
\plotone{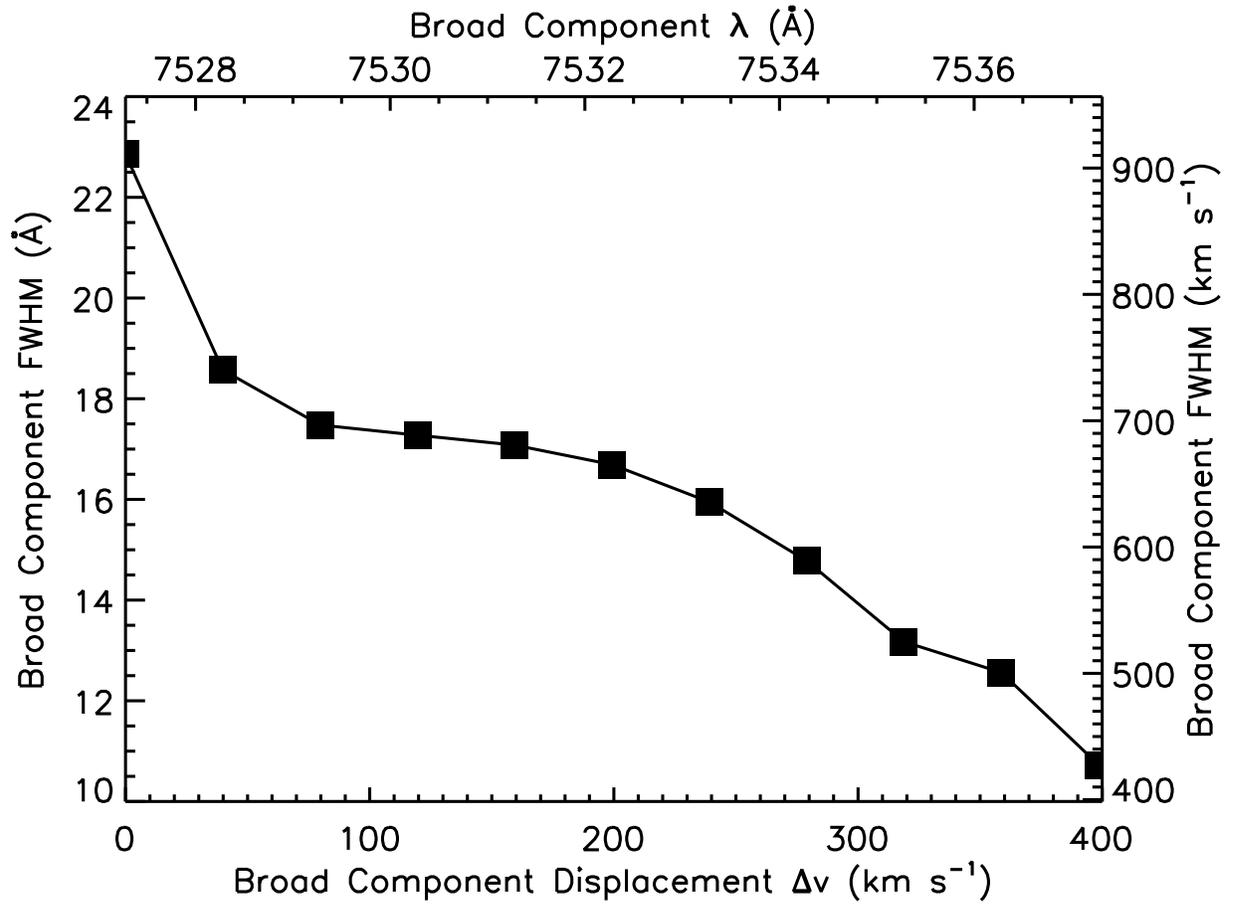}
\caption{
The relationship between the displacement velocity and the width of the
broad emission component.  In each case, the centroids for the narrow
emission component and the absorption component were fixed at the
minimum--$\chi^2$ value; the separation between the narrow emission
component and the broad component was set arbitrarily; and all other
emission parameters were left unconstrained.  When the displacement
velocity of the broad component is lowered from its best--fit value, the
width of the component increases.  Hence, there is no escaping a very
energetic component to the \lya\ emission of ES1. 
}
\label{mvred}
\end{figure}

We explored a variety of alternative kinematic scenarios for the ES1 emission,
most notably by fixing
the centroid for the central emission component and arbitrarily sliding
the separation between the central component and the broad component over
a range of values.  These models suffered from worsened $\chi^2$, with
broader emission demanded as the displacement velocity was diminished.
Figure~\ref{mvred} illustrates this trend, implying that even if the
minimum--$\chi^2$ model in Figure~\ref{modspec} over--estimates the
separation between the central component and the broad component of the
\lya\ emission, the broad component itself can only get broader.  That is,
no matter what combination of fit--parameters one chooses, there is no
escaping a very energetic component to the \lya\ emission of ES1.
In a similar vein, it is also worthy of note that models which
do not include a high--velocity compenent (e.g.\ models with
a single Gaussian emission compenent) also suffer from a
worsened $\chi^2$ compared to that of the model in Figure~\ref{modspec}.

\section{Discussion and Conclusion}
\label{discussion}

The spectral profile of the \lya\ emission of ES1 presents evidence for a
galaxy--scale outflow with a velocity of $v > 300$ km s$^{-1}$.  Of
course, as \citet{heckman00} caution, the outflow rate of a galactic wind
cannot necessarily be interpreted as the rate at which mass or energy {\it
escapes} into the IGM, since the observable manifestation of an outflow
may be produced by material still deep inside the gravitational potential
well of the galaxy halo.  Nonetheless, the outflow velocity estimated for
ES1 far exceeds the escape speed of a nominally low mass ($M < 10^{10}
M_\sun$) pregalactic fragment, consistent with the general observation
that hot gas can readily escape from dwarf galaxies, though perhaps not
from more massive systems \citep{heckman00,heckman00p,martin99}.

This conclusion bears on a host of cosmological issues surrounding the
evolution of galaxies and the IGM at high redshift.  Foremost, it suggests
that processed material from ES1 will become available to the IGM,
potentially providing the enrichment necessary to account for the amount
of metals there observed.  Indeed, recent observations of \ion{C}{4}
absorption systems along the lines--of--sight to lensed QSOs call for
enrichment at increasingly high redshift, beyond even $z > 5$ (e.g.\
Aguirre et al.\ 2001; Rauch, Sargent, \& Barlow 2001). Additionally, both
detailed observations and careful theoretical studies demand a mechanism
for pre--heating the material out of which galaxy clusters ultimately
collapse and become bound \citep[e.g.][and references
therein]{kaiser91,mushotzky97}. Here again, galaxy--scale outflows at high
redshift are the likely culprit \cite[e.g.][]{renzini93}.  Finally,
galactic winds have proved important in reproducing the faint--end slope
of the observed field galaxy luminosity function in semi--analytic models
of galaxy formation. Outflows are invoked to suppress star--formation in
low--mass dark matter halos, either via the direct escape of gas--phase
baryons in the outflow itself \citep[e.g.][]{somerville99}, or by ram
pressure stripping of the gas--phase baryons by energetic winds from
neighboring galaxies \citep{scannapieco01}.

As a somewhat speculative conclusion, we now consider the expected
correlation between strong galactic outflows and the escape of Lyman
continuum radiation from star--forming galaxies.  This correlation bears
directly on the much--debated physical nature and relative contributions
of the sources which comprise the UV background, as a significant
contribution by sources other than QSOs is required at high redshift,
owing to the rapid decline in the space density of optical and radio--loud
quasars at $z > 3$ \citep*{bianchi01,madau99}.

It is likely that star--forming galaxies fill this niche.  From the
theoretical standpoint, mechanical energy deposition in the form of
supernovae and stellar winds is expected to result in an over--pressured
cavity of hot gas inside star--forming galaxies.  In galaxies for which
the star--formation rate per unit area $\Sigma_\ast \geq 10^{-1} M_\sun \;
\mbox{yr}^{-1} \; \mbox{kpc}^{-2}$, the superbubble ultimately expands and
bursts out into the galaxy halo, allowing for the escape of hot gas and
facilitating the leak of Lyman continuum photons
\citep{heckman00,tenorio99}.  Of course, as the superbubble will expand in
the direction of the vertical pressure gradient, the burst is expected to
take the form of a weakly collimated, bipolar wind.  Hence, the leak of UV
radiation may depend sensitively on not only the distribution of neutral
gas and dust in the galaxy interstellar medium, but on the inclination of
the system. Nonetheless, from the observational standpoint,
\citet*[][hereafter SPA01]{steidel01} report the detection of significant
Lyman continuum emission in a composite spectrum of 29 Lyman--break
galaxies at $\langle z \rangle = 3.40 \pm 0.09$, suggesting an escape
fraction\footnote{Here, \fesc\ is the fraction of emitted 900 \AA\ photons
that escapes the galaxy without being absorbed by interstellar material,
normalized by the fraction of emitted 1500 \AA\ photons which similarly
escapes.  As SPA01 point out, this definition differs from definitions
encountered elsewhere, which typically consider only the fraction of
emitted 900 \AA\ photons which escapes (e.g.\ Bianchi et al.\ 2001;
Heckman et al.\ 2001; Hurwitz, Jelinsky, \& Dixon 1997; Leitherer et al.\
1995).} of UV ionizing photons of $f_{\mbox{\tiny esc}} \simgtr 0.5$.

\begin{figure}
\centering
\epsscale{1.0}
\plotone{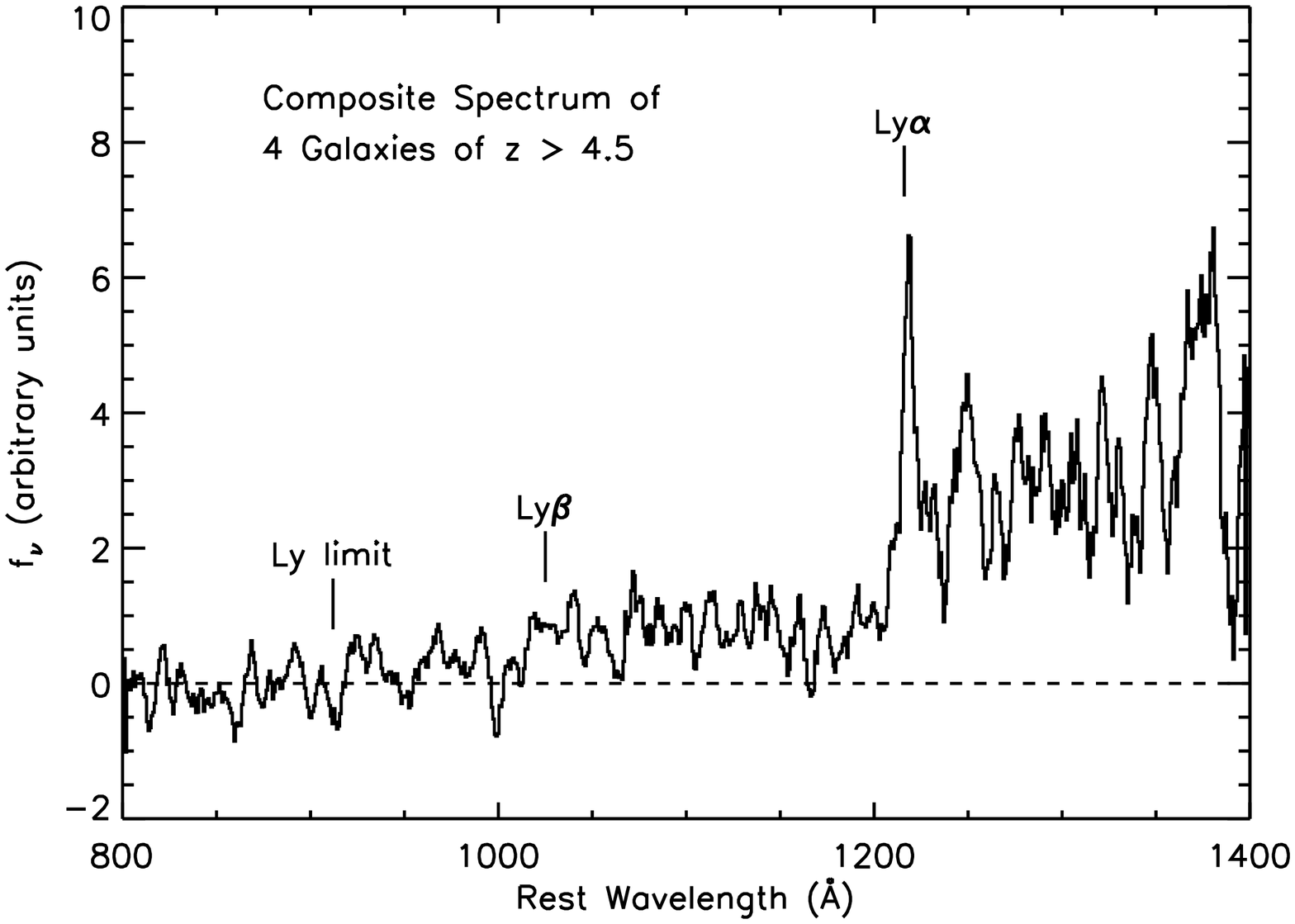}
\caption{
Composite spectrum of four galaxies at $z > 4.5$:  RD581 with $z = 4.89$,
HDF 3--951 with $z = 5.34$ \citep{spinrad98}, HDF 4--625 with $z = 4.58$,
and HDF 4--439 with $z = 4.54$ \citep[both][]{stern99}. Following
\citet{steidel01}, the composite was constructed by shifting each
flux--calibrated, one--dimensional spectrum to the rest frame, scaling to
a common median, and then combining with a simple algorithm which rejected
$1 \sigma$ outliers at each pixel. This rejection scheme removed very
nearly one point per pixel. The spectrum has been boxcar--smoothed by one
resolution element.
}
\label{composite}
\end{figure}

Given the evidence for a strong outflow in ES1, it would be intriguing to
measure the flux of photons below $\lambda = 912$ \AA.  However, ES1 is
very faint even above \lya; we estimate that it fades to $I_{\mbox{\tiny
AB}} > 29$ below the Lyman limit.  We can do only slightly better at high
redshifts with more accessible spectra: even in a composite of four of the
highest signal--to--noise Keck spectra of galaxies at $z > 4.5$ collected
by Spinrad and collaborators (Figure~\ref{composite}), our measurement of
the flux of 900 \AA\ photons is consistent with zero at the $1 \sigma$
level.  This result translates to the coarsely constrained flux ratio
$\fele / \fnin = 16.7 \pm 51.9$ ($1 \sigma$ uncertainty).  To convert this
value to the more useful $\ffif / \fnin$ ratio, we adopt an empirical
correction factor based on the set of fluxes given in SPA01, yielding an
effective $\ffif / \fnin = 31.4 \pm 98.2$. Finally, for the same intrinsic
Lyman discontinuity of 3 adopted by SPA1, we find an escape fraction of
$\fesc \simgtr 0.1 \pm 0.3$.  As we did not correct our initial $\fele /
\fnin$ ratio for the opacity of the IGM, this value represents a lower
limit. Evidently, to satisfactorily constrain the correlation between
outflow dynamics and the escape of UV ionizing photons in an individual
high redshift galaxy, we will require spectroscopy of a lensed, blue
candidate system viewed along the outflow direction.


\acknowledgments

We are humbly indebted to E. Scannapieco and J. Walters for making
generous and substantial contributions to this work, and for providing
prodigious comic relief.  We are grateful to A. Barger for making public
optical images of the HDF and its Flanking Fields of which we have made
extensive use; to C. Steidel for providing the target which resulted in
this fortuitous detection; to M. Hunt for unwittingly providing software
which aided in the reduction of the echelle data;  to M. Dickinson for
acting as the steward of the HDF and for providing the North West Flanking
Field mosaic included in this work; to the anonymous referee for
providing careful, useful commentary; and to the expert staff of the Keck
Observatory for their invaluable assistance in making the observation.  
The work of SD was supported by IGPP--LLNL University Collaborative
Research Program grant \#02--AP--015.  HS gratefully acknowledges NSF
grant AST 95--28536 for supporting much of the research presented herein.  
AD acknowledges partial support from NASA HF--01089.01--97A and from NOAO.
NOAO is operated by AURA, Inc., under cooperative agreement with the NSF.
The work of DS was carried out at the Jet Propulsion Laboratory,
California Institute of Technology, under contract with NASA.  The work of
SD, WvB, WdV, and MR was performed under the auspices of the U.S.  
Department of Energy, National Nuclear Security Administration by the
University of California, Lawrence Livermore National Laboratory under
contract No.\ W--7405--Eng--48.  This work made use of NASA's Astrophysics
Data System Abstract Service.



\eject

\eject

\begin{deluxetable}{lcccc}
\tablewidth{0pt}
\tablecolumns{5}
\tablecaption{ES1 \lya\ Line Model Parameters}
\tablehead{\colhead{Component} & \colhead{$\lambda$} & \colhead{Peak Amplitude} &
\colhead{FWHM} & \colhead{$\Delta v$\tablenotemark{\dag}} \\
\colhead{ } & \colhead{(\AA)} & \colhead{($10^{-18}$ ergs s$^{-1}$ cm$^{-2}$ \AA$^{-1}$)} &
\colhead{(\AA\ / km s$^{-1}$)} & \colhead{(km s$^{-1}$)}} 
\startdata 
Narrow Component     & 7527 & \phs 3.16 & \phm{1}7 / 280  & \phs \nodata \\
Broad Component      & 7535 & \phs 0.62 & 14 / 560        & \phs 320  \\
Absorption Component & 7518 & $-0.06$   & 20 / 800        & $-360$ \\
\enddata
\tablenotetext{\dag}{
Displacement velocity relative to the central, narrow emission component. 
Positive velocity is a redshift; negative velocity is a blueshift.}
\label{fittab}
\end{deluxetable}

\begin{deluxetable}{cc}
\tablewidth{0pt}
\tablecolumns{2}
\tablecaption{ES1 Ground--Based Photometry}
\tablehead{\colhead{Band} & \colhead{Magnitude} \\
\colhead{ } & \colhead{(Vega--based, 1\farcs5 aperture)}}
\startdata
$V$        & $> 26.1$\tablenotemark{\dag} \\
$R$        & $> 27.5$\tablenotemark{\dag} \\
$I$        & $25.1 \pm 0.1$\tablenotemark{\ddag} \\ 
$z$        & $> 25.2$\tablenotemark{\dag} \\
\enddata
\tablenotetext{\dag}{$3 \sigma$ limiting magnitude.}
\tablenotetext{\ddag}{Add $\sim 0.4$ magnitudes to this Vega--based
magnitude for comparison to the AB isophotal magnitude cited in 
\S~\ref{observation}.}
\label{phot}
\end{deluxetable}

\end{document}